\documentclass[prd,amsmath,amssymb,showpacs,superscriptaddress,nofootinbib]{revtex4}
\usepackage{amssymb}
\usepackage{graphicx,epsfig}
\usepackage{dcolumn}
\usepackage{bm}
\usepackage[dvipdfm,CJKbookmarks=true,unicode,colorlinks,linkcolor=blue,anchorcolor=blue,citecolor=blue,pdfborder={0 0 0}]{hyperref}
\usepackage{amsfonts}
\usepackage{keyval,graphicx}
\usepackage{textcomp,wasysym}

\begin{document}

\normalsize

\parskip=5pt plus 1pt minus 1pt

\title{\boldmath Study of $a^{0}_{0}$(980)--$f_{0}$(980) mixing}

\author{
{\small M.~Ablikim$^{1}$, M.~N.~Achasov$^{5}$, L.~An$^{9}$, Q.~An$^{36}$, Z.~H.~An$^{1}$, J.~Z.~Bai$^{1}$, R.~Baldini$^{17}$, Y.~Ban$^{23}$, J.~Becker$^{2}$, N.~Berger$^{1}$
% ...
\\
\vspace{0.2cm}
(BES~III Collaboration)\\
\vspace{0.2cm} {\it
$^{1}$ Institute of High Energy Physics, Beijing 100049, P. R. China\\
$^{2}$ Bochum Ruhr-University, 44780 Bochum, Germany\\
% ...
$^{5}$ G.I. Budker Institute of Nuclear Physics SB RAS (BINP), Novosibirsk 630090, Russia\\
% ...
$^{9}$ Guangxi University, Naning 530004, P. R. China\\
% ...
$^{17}$ Laboratori Nazionali di Frascati - INFN, 00044 Frascati, Italy\\
% ...
$^{23}$ Peking University, Beijing 100871, P. R. China\\
% ...
$^{36}$ University of Science and Technology of China, Hefei 230026, P. R. China\\
% ...
}}
\vspace{0.4cm} }

\begin{abstract}

Using samples of $2.25 \times 10^{8}$ $J/\psi$ events and $1.06 \times
10^{8}$ $\psi^{\prime}$ events collected with the BES~III detector, we
study the $f_{0}(980)\to a^{0}_{0}(980)$ and $a^{0}_{0}(980)\to
f_{0}(980)$ transitions in the processes $J/\psi\to\phi
f_{0}(980)\to\phi a^{0}_{0}(980)$ and $\chi_{c1}\to\pi^{0}
a^{0}_{0}(980)\to\pi^{0} f_{0}(980)$, respectively. Evidence for
$f_{0}(980)\to a^{0}_{0}(980)$ is found with a significance of
3.4~$\sigma$, while in the case of $a^{0}_{0}(980)\to f_{0}(980)$
transition, the significance is 1.9~$\sigma$. Measurements and upper
limits of both branching ratios and mixing intensities are determined.

\end{abstract}

\pacs{25.75.Gz, 14.40.Df, 12.38.Mh}

\maketitle

\section{Introduction}

The nature of the scalar mesons $a_0(980)$ and $f_0(980)$ has been a
hot topic in light hadron physics for many years. These two states,
with similar masses but different decay modes, are difficult to
accommodate in the constituent quark-antiquark scenario. Tremendous
efforts in both experiment and theory have been made in order to
understand them. Suggestions for their being exotic candidates, such as
tetra-quark states, hybrids or $K\bar{K}$ molecules, can be found in
the literature~\cite{Jaffe:1976ig,Achasov:1987ts,Achasov:1997ih,
Weinstein:1983gd,Weinstein:1990gu,SI_conf1995}.

The possibility of mixing between $a^{0}_{0}(980)$ and $f_{0}(980)$
was suggested long ago, and its measurement will shed light on the
nature of these two
resonances~\cite{Achasov:1979xc,Hanhart:2007bd,Achasov:2002hg,
Kerbikov:2000pu,Achasov:2003se,Close:2000ah,Wu:2007jh,Wu:2008hx}. In
particular, the leading contribution to the isospin-violating mixing
transition amplitudes for $f_{0}(980)\to a^{0}_{0}(980)$ and
$a^{0}_{0}(980)\to f_{0}(980)$, is shown to be dominated by the
difference of the unitarity cut which arises from the mass difference
between the charged and neutral $K\bar{K}$ pairs. As a consequence, a
narrow peak of about 8 MeV$/c^2$ is predicted between the charged and
neutral $K\bar{K}$
thresholds~\cite{Hanhart:2007bd,Wu:2007jh,Wu:2008hx}.

The mixing amplitudes strongly depend on the couplings of $a_0^0(980)$
and $f_0(980)$ to $K\bar{K}$, and to probe the properties of these two
scalar states, precise measurements of the mixing transitions are very
important. Two kinds of mixing intensities, i.e. $\xi_{fa}$ and
$\xi_{af}$ for the $f_{0}(980)\to a^{0}_{0}(980)$ and
$a^{0}_{0}(980)\to f_{0}(980)$ transitions, respectively, can be
defined and are accessible to measurement in charmonium
decays~\cite{Wu:2007jh,Wu:2008hx}:
\begin{eqnarray*}\label{eq_xifa}
\xi_{fa} \equiv \frac {Br(J/\psi\to\phi f_{0}(980)\to\phi
a^{0}_{0}(980)\to\phi\eta\pi^{0})} {Br(J/\psi\to\phi
f_{0}(980)\to\phi\pi\pi)},
\end{eqnarray*}
and
\begin{eqnarray*}\label{eq_xiaf}
\xi_{af} \equiv \frac {Br(\chi_{c1}\to\pi^{0}
a^{0}_{0}(980)\to\pi^{0} f_{0}(980)\to\pi^{0}\pi^{+}\pi^{-})}
{Br(\chi_{c1}\to\pi^{0} a^{0}_{0}(980)\to\pi^{0}\pi^{0}\eta)}.
\end{eqnarray*}

Using samples of $2.25 \times 10^{8}$ $J/\psi$
events~\cite{yanghx} and $1.06 \times 10^{8}$ $\psi^{\prime}$
events~\cite{Ablikim:2010zn} collected with the BES~III detector in
2009, we perform direct measurements of the
$a^{0}_{0}(980)-f_{0}(980)$ mixing intensities via the processes
$J/\psi\to\phi f_{0}(980)\to\phi a^{0}_{0}(980)\to\phi\eta\pi^{0}$
and $\chi_{c1}\to\pi^{0} a^{0}_{0}(980)\to\pi^{0}
f_{0}(980)\to\pi^{0}\pi^{+}\pi^{-}$.

\section{Detector and Monte Carlo simulation}

BEPC~II is a double-ring $e^{+}e^{-}$ collider designed to provide a
peak luminosity of $10^{33}$ cm$^{-2}s^{-1}$ at a beam current of
$0.93$~A. The BES~III detector has a geometrical acceptance of
$93\%$ of $4\pi$ and has four main components: (1) A small-cell,
helium-based ($40\%$ He, $60\%$ C$_{3}$H$_{8}$) main drift chamber
(MDC) with $43$ layers providing an average single-hit resolution of
$135$~$\mu$m, charged-particle momentum resolution in a $1$~T
magnetic field of $0.5\%$ at 1~GeV$/c$, and the $dE/dx$ resolution
that is better than $6\%$. (2) An electromagnetic calorimeter (EMC)
consisting of $6240$ CsI(Tl) crystals in a cylindrical structure
(barrel) and two endcaps. The energy resolution at $1.0$~GeV$/c$ is
$2.5\%$ ($5\%$) in the barrel (endcaps), and the position resolution
is $6$~mm ($9$~mm) in the barrel (endcaps). (3) A time-of-flight
system (TOF) constructed of $5$-cm-thick plastic scintillators, with
$176$ detectors of $2.4$~m length in two layers in the barrel and
$96$ fan-shaped detectors in the endcaps. The barrel (endcap) time
resolution of $80$~ps ($110$~ps) provides $2\sigma$ $K/\pi$
separation for momenta up to $\sim 1.0$~GeV$/c$. (4) The muon system
(MUC) consists of $1000$~m$^{2}$ of Resistive Plate Chambers (RPCs)
in nine barrel and eight endcap layers and provides $2$~cm position
resolution.

The efficiency for $J/\psi\to\phi f_{0}(980)\to\phi
a^{0}_{0}(980)\to\phi\eta\pi^{0}$ is estimated using a Monte Carlo (MC)
simulation of $J/\psi\to\phi S$, where $S$ is the mixing signal
represented by a narrow scalar Breit-Wigner uniformly decaying to $\eta\pi^{0}$.
The mass of the mixing signal is set to be $991.3$~MeV$/c^{2}$
(the center of $(m_{K^{+}} + m_{K^{-}})$ and $(m_{K^{0}} +
m_{\bar{K}^{0}})$~\cite{PDG}) and the width of the mixing
signal is set to be 8~MeV$/c^{2}$.  The efficiency for $\psi^{\prime}\to\gamma\chi_{c1}
\to\gamma\pi^{0} a^{0}_{0}(980)\to\gamma\pi^{0} f_{0}(980)\to\gamma\pi^{0}\pi^{+}\pi^{-}$ is
estimated using a Monte Carlo simulation of $\psi^{\prime}\to\gamma \chi_{c1}\to\gamma\pi^{0} S$,
where $S$ is the mixing signal with parameters
as above, and decays into $\pi^{+}\pi^{-}$ isotropically.

\section{Event selection}

Tracks of charged particles in BES~III are reconstructed from MDC
hits. We select tracks within $\pm 20$~cm of the interaction point
in the beam direction and within $2$~cm in the plane perpendicular
to the beam. The TOF and $dE/dx$ information are combined to form
particle identification (PID) confidence levels for the $\pi$, $K$,
$p$ hypotheses; each track is assigned to the particle type that
corresponds to the hypothesis with the highest confidence level.

Photon candidates are reconstructed by clustering EMC crystal
energies. Efficiency and energy resolution are improved by including
energy deposits in nearby TOF counters. The minimum energy is $25$~MeV
for barrel showers ($|\cos\theta| < 0.80$) and $50$~MeV for endcap
showers ($0.86<|\cos\theta|<0.92$). To exclude showers from charged
particles, the angle between the nearest charged track and the shower
must be greater than $10^{\circ}$. EMC cluster timing requirements
suppress electronic noise and energy deposits unrelated to the event.

The $\pi^{0}\to\gamma\gamma$ and $\eta\to\gamma\gamma$ candidates are
formed from pairs of photon candidates that are kinematically fit to
the known resonance masses, and the $\chi^{2}$ from the kinematic fit
with one degree of freedom is required to be less than $25$. The decay
angle of a photon is the polar angle measured in the $\eta$ or
$\pi^{0}$ rest frame with respect to the $\eta$ or $\pi^{0}$ direction
in the $J/\psi$ or $\psi^{\prime}$ rest frame. Real $\eta$ and
$\pi^{0}$ mesons decay isotropically, and their angular distributions
are flat. However, the $\eta$ and $\pi^{0}$ candidates which originate
from a wrong photon combination do not have a flat distribution in
this variable. To remove wrong photon combinations, the decay angle is
required to satisfy $|\cos\theta_{decay}|<0.95$.

\section{Measurement of $J/\psi\to\phi f_{0}(980)\to\phi
a^{0}_{0}(980)\to\phi\eta\pi^{0}$} \label{jpsi_channel}

\begin{figure}[htbp]
   \centerline{
   \psfig{file=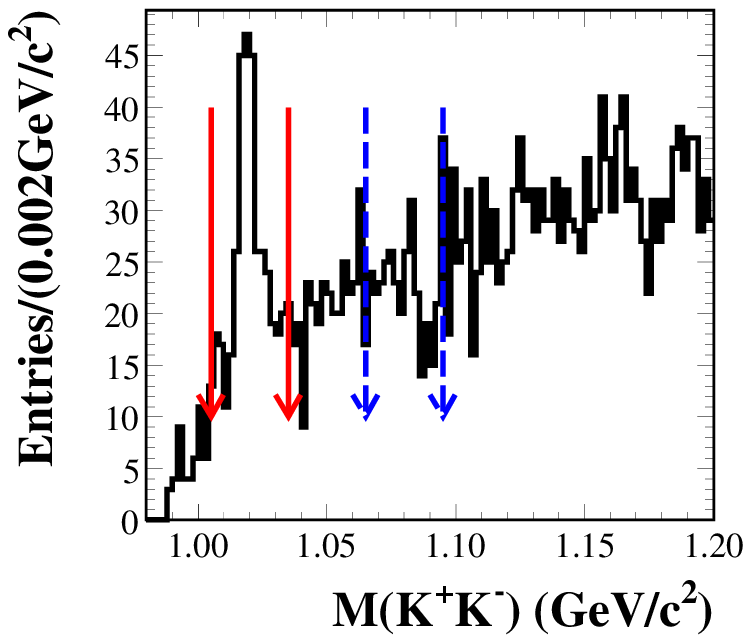,width=8cm, angle=0}
              \put(-170,170){(a)}
   \psfig{file=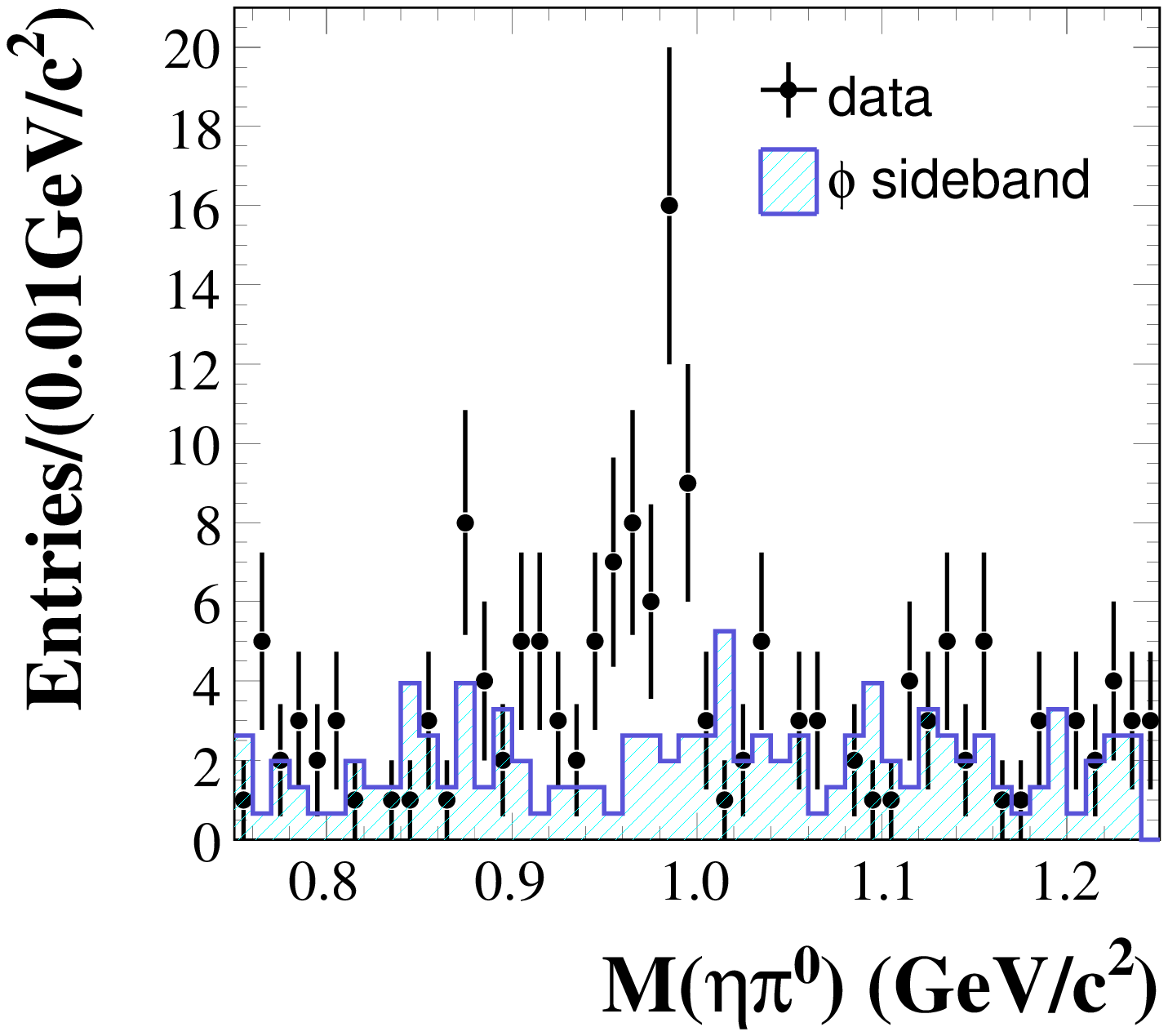,width=8cm, angle=0}
              \put(-170,170){(b)}}
   \centerline{
   \psfig{file=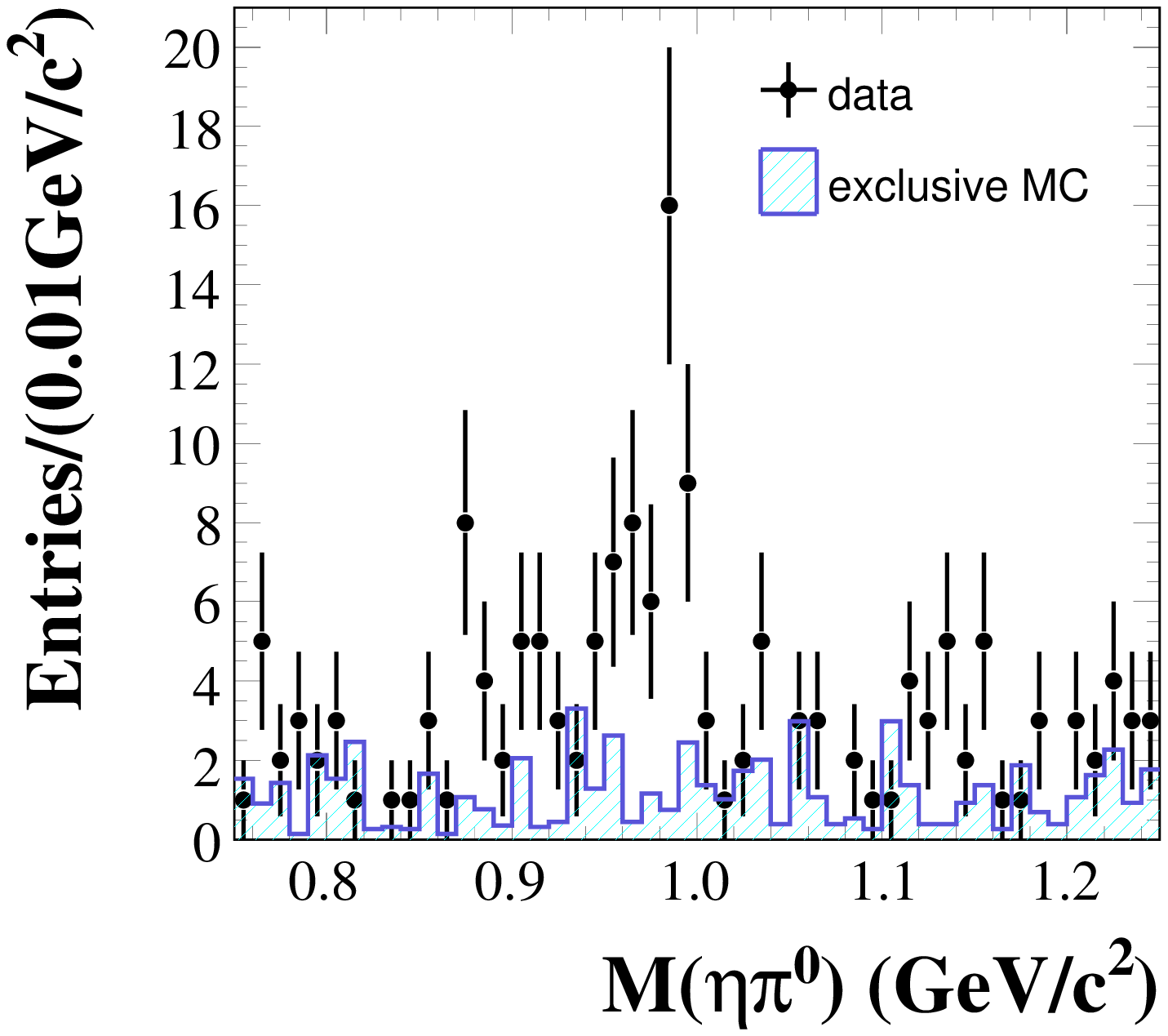,width=8cm, angle=0}
              \put(-170,170){(c)}
   \psfig{file=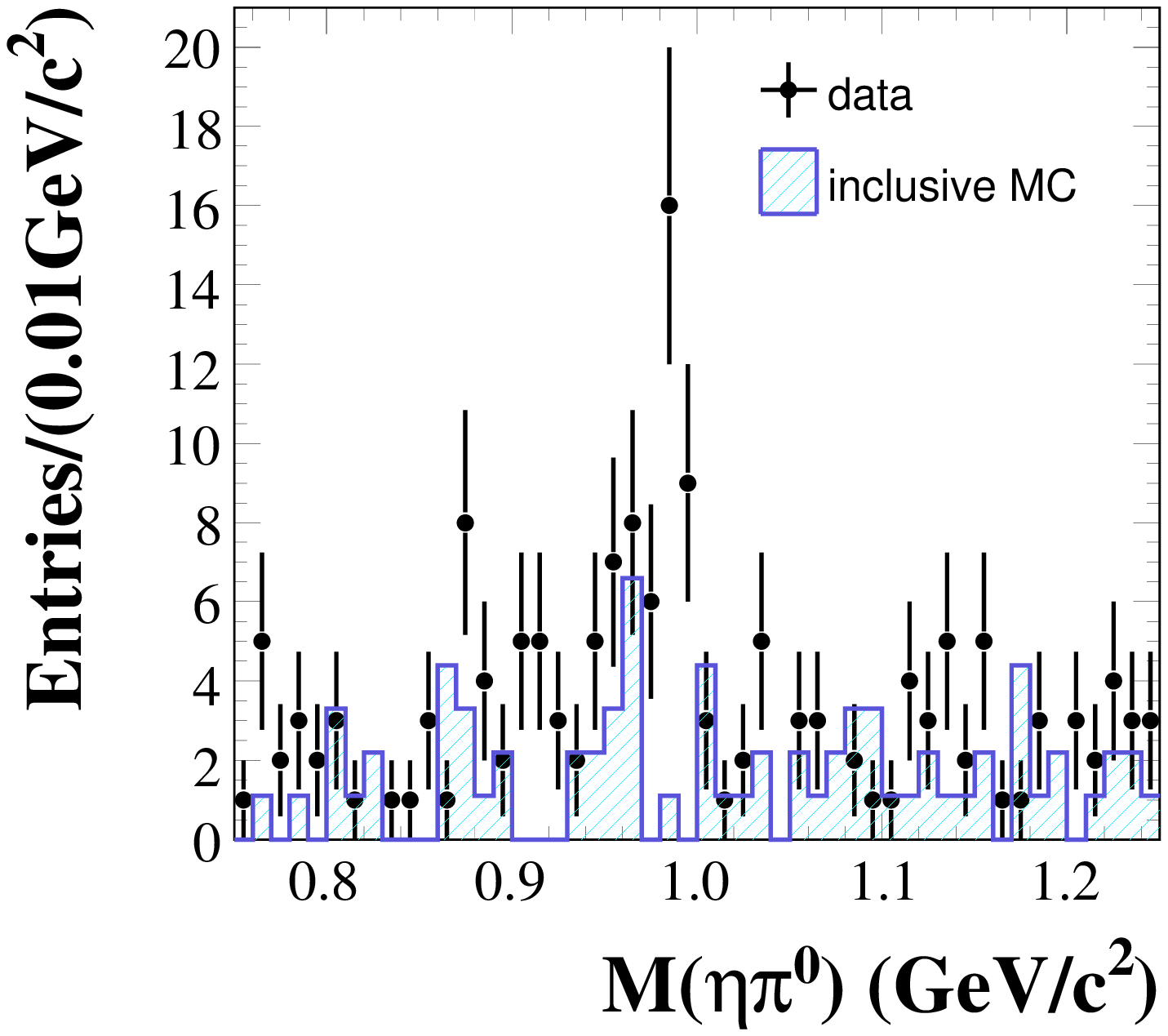,width=8cm, angle=0}
              \put(-170,170){(d)}}
   {\caption{(a) The invariant mass spectrum of $K^{+}K^{-}$
   in $J/\psi\to K^{+}K^{-}\eta\pi^{0}$.
   The solid arrows show the $\phi$ mass window.
   The dashed arrows show the $\phi$ sideband region used to
   estimate backgrounds. The $\eta\pi^{0}$ invariant mass of selected events
   is shown in (b), (c) and (d). The dots with error bars show the mass spectrum
   of $M_{\eta\pi^{0}}$ recoiling against the $\phi$. The shaded histogram is:
   (b) recoiling against the $\phi$ sideband; (c) from exclusive MC; (d)
   from inclusive MC.}
   \label{jpsi_plot1}}
\end{figure}

Events with two oppositely charged tracks identified as kaons and
at least one distinct $\pi^{0}$ candidate and $\eta$ candidate are selected.
A six-constraint kinematic fit (6C) is performed to the $J/\psi\to
K^{+}K^{-}\eta\pi^{0}$ hypothesis (constraints are the 4-momentum of the
$J/\psi$ and the $\pi^{0}$ and $\eta$ masses), and $\chi^{2}_{6C}<60$ is
required. If there is more than one combination,
the combination with the smallest $\chi^{2}_{6C}$ is retained.
The events are fitted to $J/\psi\to K^{+}K^{-}\eta\eta$ and $J/\psi\to
K^{+}K^{-}\pi^{0}\pi^{0}$, and the probabilities are required to be
less than that from the kinematic fit to the signal channel
$J/\psi\to K^{+}K^{-}\eta\pi^{0}$. The $K^{+}K^{-}$ invariant mass
distribution of selected events is shown in Fig.~\ref{jpsi_plot1} (a),
where a clear $\phi$ signal can be seen. The solid and dashed arrows show
the signal and sideband regions, respectively.

Figure~\ref{jpsi_plot1} (b) shows the $\eta\pi^{0}$ invariant mass
distribution recoiling against the $\phi$ signal
($|m_{K^{+}K^{-}}-1.02|<0.015$~GeV$/c^{2}$). A narrow structure appears
around $980$~MeV$/c^{2}$. The shaded histogram shows the $\eta\pi^{0}$
invariant mass of events recoiling against the $\phi$ sideband
($1.065$~GeV$/c^{2}<M_{K^{+}K^{-}}<1.095$~GeV$/c^{2}$), and no sign of a peak near
$980$~MeV$/c^{2}$ is evident.

Exclusive MC samples of $J/\psi$ decays which have similar final
states are generated to check whether a peak near $980$~MeV$/c^{2}$ can be
produced in the $\eta\pi^{0}$ mass spectrum. The main backgrounds
come from: (1) $J/\psi\to\phi\pi^{0}\pi^{0}$ via $f_{0}(980)$,
$f_{2}(1270)$ or a phase space process; and (2) $J/\psi\to K^{*}\bar{K}\eta+c.c.$,
$J/\psi\to K^{*}_{2}(1430)\bar{K}\eta+c.c.$, $J/\psi\to\gamma
f_{2}(1950)\to\gamma K^{*}\bar{K}^{*}$. The $\eta\pi^{0}$
invariant mass distribution from all these possible background
channels is smooth, shown as the shaded region in Fig.~\ref{jpsi_plot1} (c),
and will not affect the determination of the number of mixing events.

A MC sample of $2.0\times 10^{8}$ inclusive $J/\psi$ decay events is
used to investigate other possible backgrounds too. The shaded
histogram in Fig.~\ref{jpsi_plot1} (d) shows the $\eta\pi^{0}$
invariant mass distributions of events selected from the inclusive
MC sample. In the $980$~MeV$/c^{2}$ region, there is no peaking
background.

An underlying process is from $J/\psi\to\phi a^{0}_{0}(980)$ via a $\gamma^{*}$ or
$K^{*}K$ loops~\cite{Wu:2007jh}. However, it will produce a much broader
distribution ($50-100$~MeV$/c^{2}$) in the $\eta\pi^{0}$ mass spectrum than
$f_{0}(980)\to a^{0}_{0}(980)$ mixing~\cite{Wu:2007jh}. We estimate its
contribution in the fit to the mass spectrum.

\begin{figure}[htbp]
   \centerline{ \psfig{file=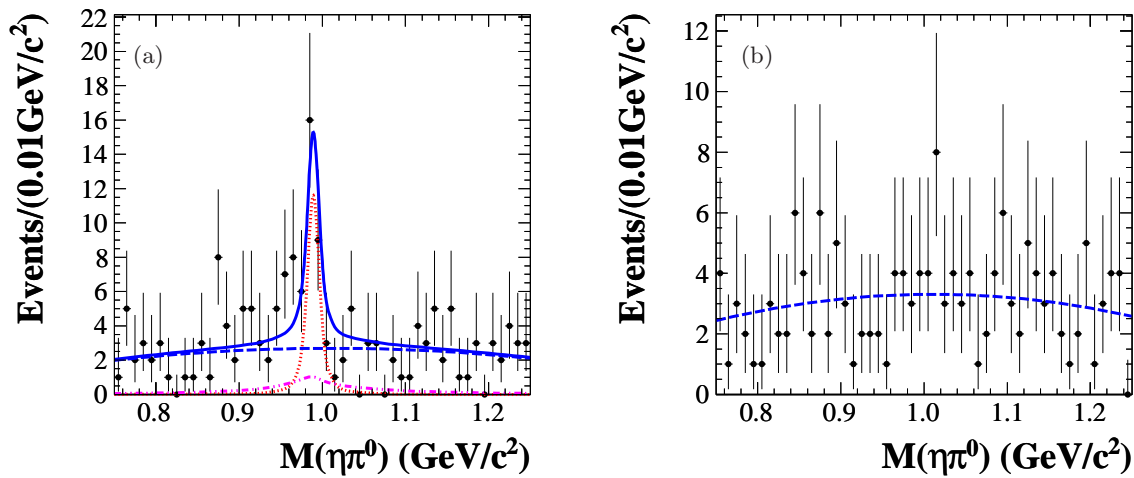,width=16cm, angle=0}
   \put(-400,170){(a)} \put(-170,170){(b)}} {\caption{ Results of the
   simultaneous fit of the $\eta\pi^{0}$ mass spectra: (a) recoiling
   against the $\phi$, (b) recoiling against the $\phi$
   sideband.  The solid curve in the result of the fit described in the
   text. The dotted curve is the mixing signal, and the dash-dotted curve
   is the $a^{0}_{0}(980)$ contribution from $\gamma^{*}$ or $K^{*}K$
   loops. The dashed curves in (a) and (b) denote the background
   polynomial.}
   \label{jpsi_fit}}
\end{figure}

A simultaneous unbinned maximum likelihood fit to the $\eta\pi^{0}$
mass spectrums recoiling against the $\phi$ and the $\phi$ sideband is performed.
In the signal region, the probability density
function (PDF) is composed of the mixing signal, represented by the
shape extracted from MC simulation of a narrow
Breit-Wigner, the $a^{0}_{0}(980)$ contribution from $\gamma^{*}$ or
$K^{*}K$ loops represented by a Flatt\'{e} formula,\footnote{The
Flatt\'{e} formula is taken from Ref.~\cite{Flatte:1976xu}. The values of
coupling constants and mass as used here are taken from the Crystal Barrel
experiment results in Ref.~\cite{Bugg:1994mg} and Ref.~\cite{Wu:2007jh}.}
and a 2nd order polynomial for the backgrounds.
In the sideband region, the PDF is a 2nd order polynomial only. The
mass of the mixing signal is set to $991.3$~MeV$/c^{2}$ (the
center of ($m_{K^{+}}+m_{K^{-}})$ and
$(m_{K^{0}}+m_{\bar{K}^{0}})$~\cite{PDG}) and the width of the
mixing signal is set to 8~MeV$/c^{2}$.  The shape parameters of the
background polynomials in the signal region and the sideband region
are constrained to vary simultaneously in the fit. The normalization
of each component is allowed to float. Figures~\ref{jpsi_fit} (a) and
(b) show the results of the simultaneous fit.

The fit yields $N(f_{0}\to a^{0}_{0}) = 25.8\pm8.6$ events for the
mixing signal, and $N(a^{0}_{0}(980)) = 13.6\pm 24.8$ events for the
$a^{0}_{0}(980)$ contribution from $\gamma^{*}$ or $K^{*}K$ loops.
Comparing with the fit result without the mixing signal, the change
in $-lnL$ with $\Delta(d.o.f.)=1$ is $5.90$, corresponding to a
statistical significance of $3.4\sigma$. Comparing with the fit
result without the $a^{0}_{0}(980)$ contribution from $\gamma^{*}$
or $K^{*}K$ loops, the change in $-lnL$ with $\Delta(d.o.f.)=1$ is
$0.14$, corresponding to a statistical significance of $0.5\sigma$.
Using the Bayesian method, the upper limit for the number of
mixing events is $37.5$, and the upper limit of the number of
$a^{0}_{0}(980)$ from $\gamma^{*}$ or $K^{*}K$ loops is $50.8$
events at the $90\%$ confidence level (C.L.). The results are listed
in Table~\ref{table_jpsi_fit}.

The branching ratio of the mixing signal $J/\psi\to\phi
f_{0}(980)\to\phi a^{0}_{0}(980)\to\phi\eta\pi^{0}$ is calculated as
:
\begin{eqnarray*}
Br(J/\psi\to\phi f_{0}(980)\to\phi
a^{0}_{0}(980)\to\phi\eta\pi^{0})=\frac {N(f_{0}\to a^{0}_{0})}
{\varepsilon_{fa} \cdot N_{J/\psi} \cdot Br(\phi\to K^{+}K^{-})
\cdot Br(\eta\to\gamma\gamma) \cdot Br(\pi^{0}\to\gamma\gamma)},
\end{eqnarray*}
where $N_{J/\psi}$ is the total number of $J/\psi$ events and
$\varepsilon_{fa}=(18.5\pm0.2)\%$ is the efficiency for the mixing
signal $J/\psi\to\phi f_{0}(980)\to\phi
a^{0}_{0}(980)\to\phi\eta\pi^{0}$. The branching ratio is then
determined to be $(3.3\pm1.1)\times10^{-6}$, where the error is
statistical only.

The total branching ratio of $J/\psi\to\phi
a^{0}_{0}(980)\to\phi\eta\pi^{0}$ is calculated as :
\begin{eqnarray*}
Br(J/\psi\to\phi\eta\pi^{0})= \frac {N(f_{0}\to
a^{0}_{0})/\varepsilon_{fa}+N(a^{0}_{0})/\varepsilon_{a} }
{N_{J/\psi} \cdot Br(\phi\to K^{+}K^{-}) \cdot
Br(\eta\to\gamma\gamma) \cdot Br(\pi^{0}\to\gamma\gamma)},
\end{eqnarray*}
where $\varepsilon_{a}=(18.3\pm0.2)\%$ is the efficiency for the
underlying process $J/\psi\to\phi a^{0}_{0}(980)\to\phi\eta\pi^{0}$.
The branching ratio is then determined to be
$(5.0\pm2.7)\times10^{-6}$, where the error is statistical only.

\begin {table}[htp]
\begin {center}
\caption {Results of fits to the mass spectrum of $\eta\pi^{0}$. The
fit is described in the text, and the yield error is statistical
only. N($f_{0}\to a^{0}_{0}$) is the number of mixing events
$f_{0}(980)\to a^{0}_{0}(980)$. S($f_{0}\to a^{0}_{0}$) is the
significance of the mixing signal $f_{0}(980)\to a^{0}_{0}(980)$.
N($a^{0}_{0}(980)$) is the number of $a^{0}_{0}(980)$ events
from $\gamma^{*}$ or $K^{*}K$ loops. S($a^{0}_{0}(980)$) is the
significance of the $a^{0}_{0}(980)$ contribution from $\gamma^{*}$ or
$K^{*}K$ loops.} \label{table_jpsi_fit} \vspace{0.2cm}
\begin {tabular}{c|c|c|c|c} \hline
     Fitting with                               &N($f_{0}\to a^{0}_{0}$) &S($f_{0}\to a^{0}_{0}$)          &N($a^{0}_{0}(980)$)         &S($a^{0}_{0}(980)$)    \\ \hline  \hline
     mixing + $a^{0}_{0}(980)$ + 2nd poly.      &$25.8\pm8.6(<37.5)$     &$3.4\sigma$                      &$13.6\pm 24.8(<50.8)$       &$0.5\sigma$            \\ \hline
     mixing + 2nd poly.                         &$28.6\pm7.0(<39.1)$     &$5.8\sigma$                      &$-$                         &$-$                    \\ \hline
     $a^{0}_{0}(980)$ + 2nd poly.               &$-$                     &$-$                              &$75.8\pm 17.3(<99.1)$       &$4.7\sigma$            \\  \hline
\end {tabular}
\end {center}
\end {table}

If we fit the $\eta\pi^{0}$ invariant mass spectrum only with the
mixing signal plus a 2nd order polynomial background, the fit yields
$28.6\pm 7.0$ events for the mixing signal. Comparing with the fit
result with only the 2nd order polynomial, the change in $-lnL$ with
$\Delta(d.o.f.)=1$ is $16.65$, corresponding to a statistical
significance of $5.8\sigma$. The upper limit at the $90\%$ C.L. is
$39.1$ events.

If we assume there is no mixing and fit the $\eta\pi^{0}$ invariant
mass spectrum only with the $a^{0}_{0}(980)$ contribution from
$\gamma^{*}$ or $K^{*}K$ loops plus a 2nd order polynomial, the fit
yields $75.8\pm 17.3$ events for the $a^{0}_{0}(980)$ contribution
from $\gamma^{*}$ or $K^{*}K$ loops. Comparing with the fit result
with only the 2nd order polynomial, the change in $-lnL$ with
$\Delta(d.o.f.)=1$ is $10.89$, corresponding to a statistical
significance of $4.7\sigma$. The upper limit at the $90\%$ C.L. is
$99.1$ events. The results are listed in Table~\ref{table_jpsi_fit}.
The total branching ratio of $J/\psi\to\phi
a^{0}_{0}(980)\to\phi\eta\pi^{0}$ is calculated to be
$(9.7\pm2.2)\times10^{-6}$, where the error is statistical only.

\section{Measurement of  $\psi^{\prime}\to\gamma\chi_{c1}\to\gamma\pi^{0} a^{0}_{0}(980)
\to\gamma\pi^{0} f_{0}(980)\to\gamma\pi^{0}\pi^{+}\pi^{-}$}
\label{psip_channel}

\begin{figure}[htbp]
   \centerline{
   \psfig{file=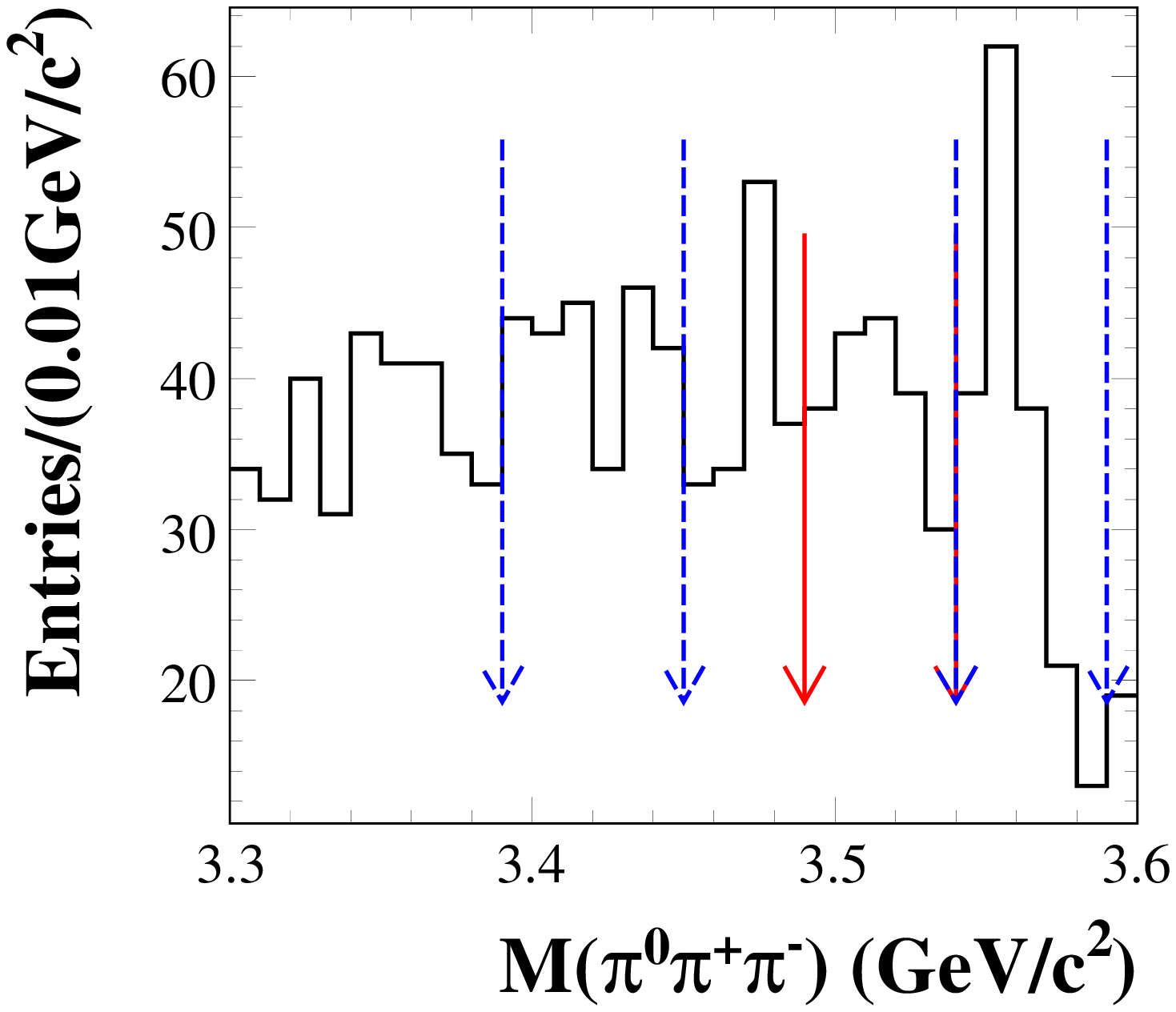,width=8cm, angle=0}
              \put(-170,170){(a)}
   \psfig{file=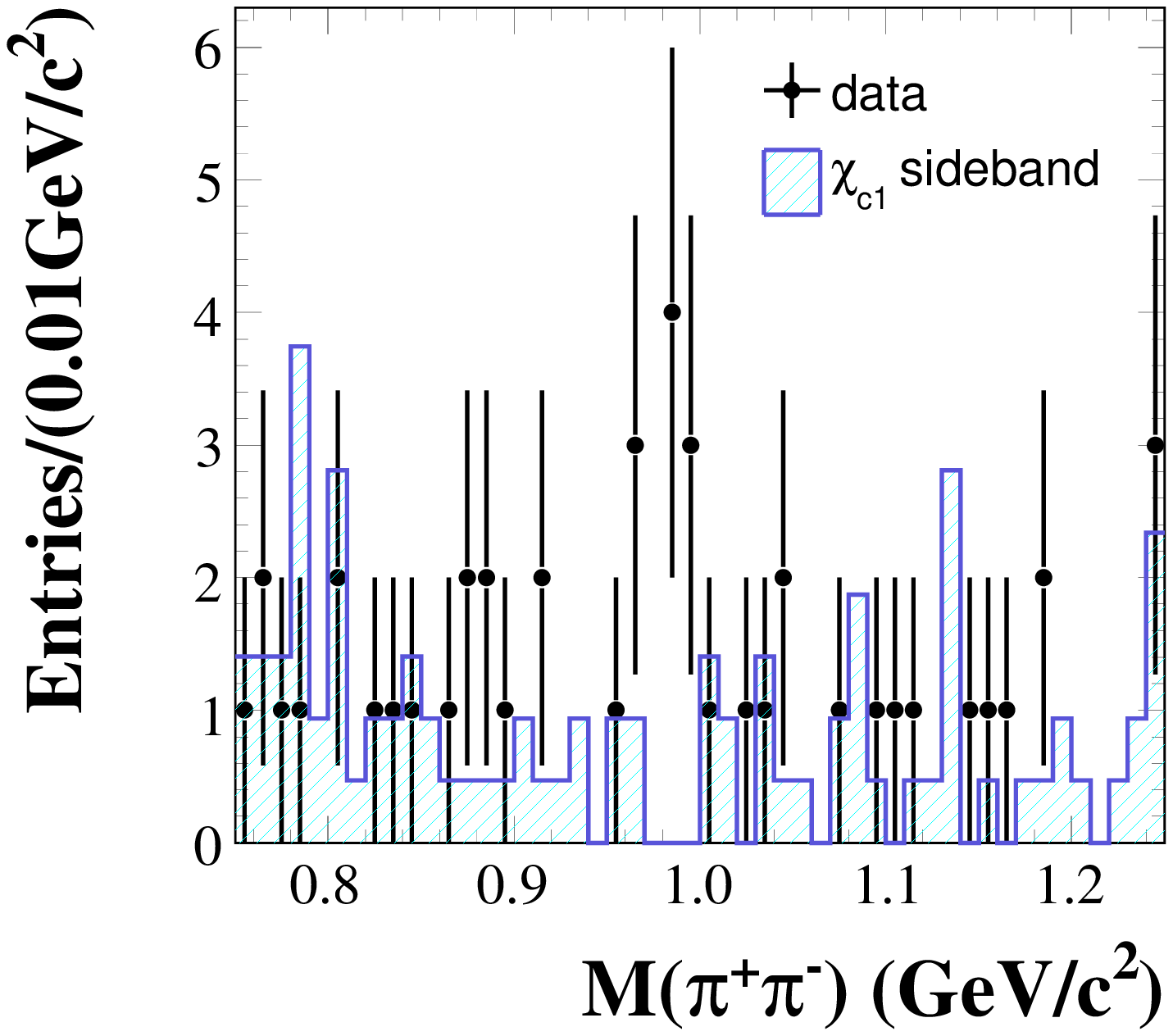,width=8cm, angle=0}
              \put(-170,170){(b)}}
   \centerline{
   \psfig{file=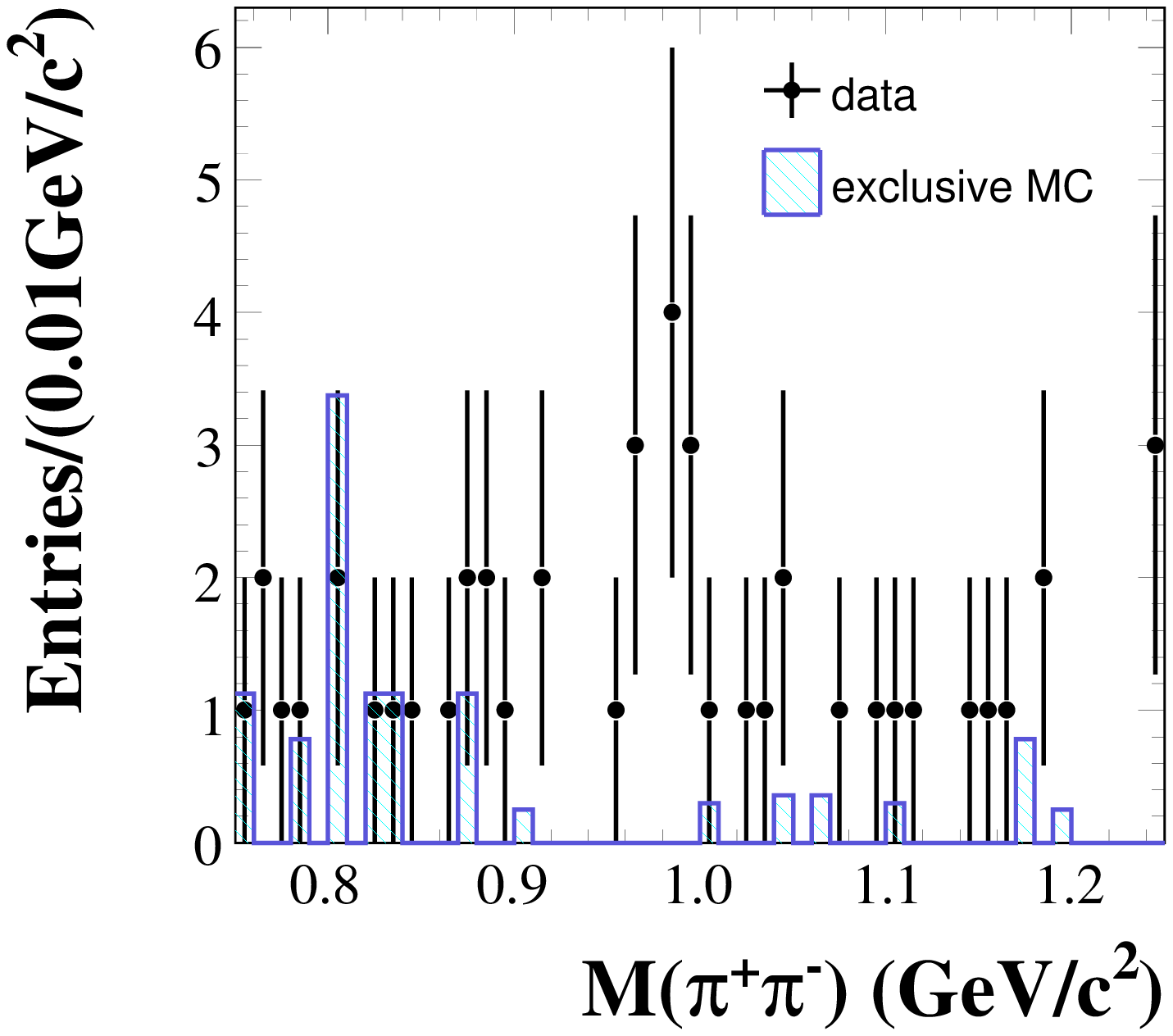,width=8cm, angle=0}
              \put(-170,170){(c)}
   \psfig{file=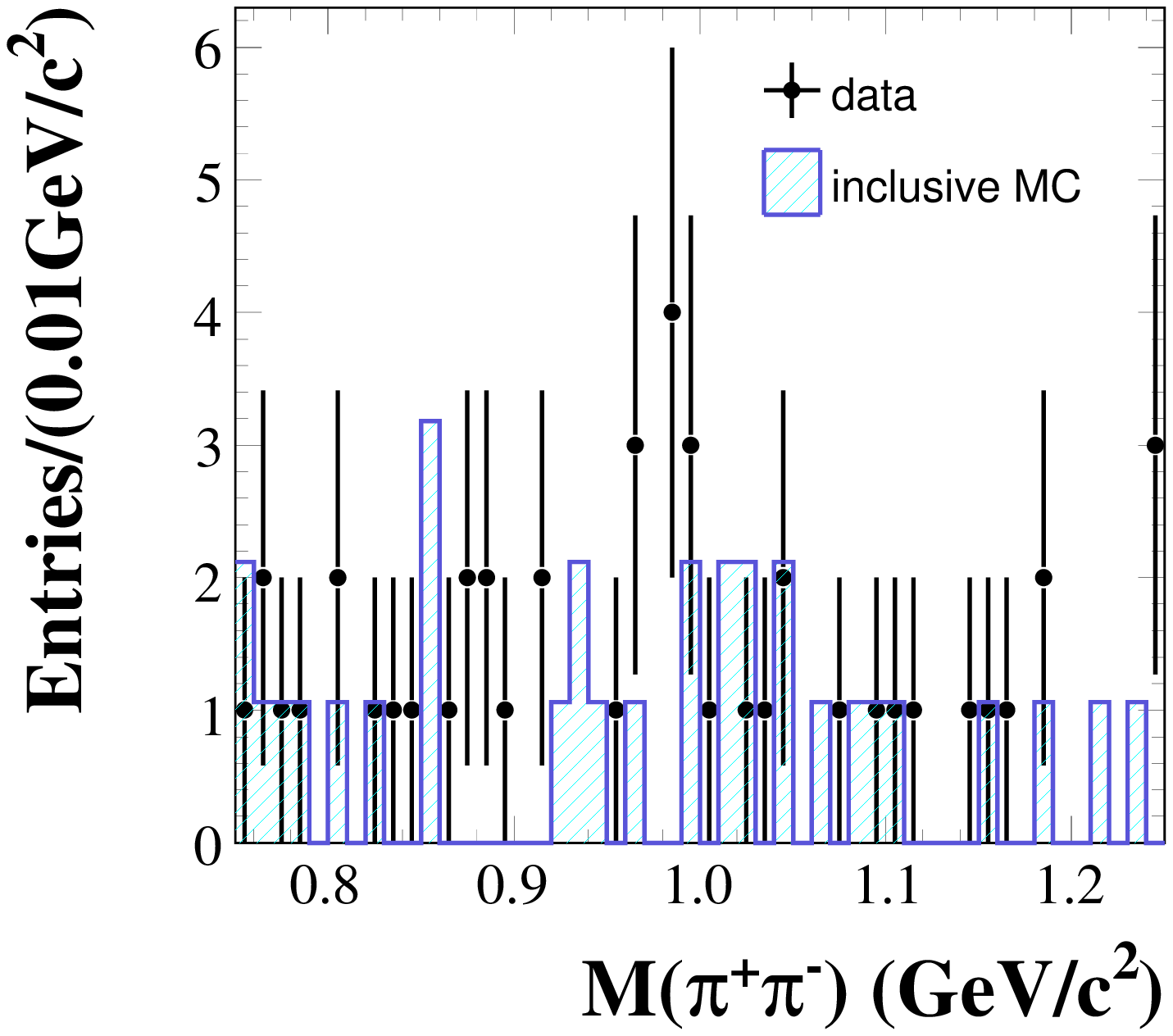,width=8cm, angle=0}
              \put(-170,170){(d)}}
   {\caption{(a) The invariant mass spectrum of $\pi^{0}\pi^{+}\pi^{-}$
   in $\psi^{\prime}\to\gamma\pi^{+}\pi^{-}\pi^{0}$.
   The solid arrows show $\chi_{c1}$ mass window, and
   the dashed arrows show the $\chi_{c1}$ sideband region used to
   estimate background. The invariant mass of $\pi^{+}\pi^{-}$ of selected
   events is shown in (b), (c) and (d). The dots with error bars show the mass
   spectrum of $M_{\pi^{+}\pi^{-}}$ in the $\chi_{c1}$ mass window.
   The shaded histogram is: (b) $\chi_{c1}$ sideband; (c) from exclusive MC;
   and (d) from inclusive MC.}
   \label{psip_plot1}}
\end{figure}

Events with two oppositely charged tracks identified as pions and at
least three photons, which form at least one distinct $\pi^{0}$
candidate, are selected.  A 5C kinematic fit is performed to the
$\psi^{\prime}\to\gamma\pi^{0}\pi^{+}\pi^{-}$ hypothesis (constraints
are the 4-momentum of $\psi^{\prime}$ and the $\pi^{0}$ mass) and
$\chi^{2}_{5C}<60$ is required. If there is more than one combination,
the combination with the smallest $\chi^{2}_{5C}$ is
retained. The events are also fitted to
$\psi^{\prime}\to\pi^{0}\pi^{+}\pi^{-}$ and
$\psi^{\prime}\to\pi^{0}\pi^{0}\pi^{+}\pi^{-}$, and the probabilities
are required to be less than that from the kinematic fit to the signal
channel $\psi^{\prime}\to\gamma\pi^{0}\pi^{+}\pi^{-}$. To remove
backgrounds with a $J/\psi$ decaying to leptons, the angle between the
two charged tracks is required to be less than $160^{\circ}$. We
further require the ratio of energy deposited by each charged track in
the EMC to its momentum measured in the MDC to be less than $0.85$.
To remove the backgrounds which have $\gamma\gamma J/\psi$ final
states, the mass recoiling from any photon pair must not be in the
$J/\psi$ mass window ($|M^{\gamma\gamma}_{
recoiling}-3.097$~GeV$/c^{2}|>0.06$~GeV$/c^{2}$).  The invariant mass
distribution of $\pi^{0}\pi^{+}\pi^{-}$ of the selected events is
shown in Fig.~\ref{psip_plot1} (a).

The $\pi^{+}\pi^{-}$ invariant mass distribution in the $\chi_{c1}$
mass window ($3.49$~GeV$/c^{2}<M_{\pi^{0}\pi^{+}\pi^{-}}<3.54$~GeV$/c^{2}$)
is shown in Fig.~\ref{psip_plot1} (b). A narrow structure
around $980$~MeV$/c^{2}$ is evident. The shaded histogram shows the
$\pi^{+}\pi^{-}$ invariant mass of events in the $\chi_{c1}$
sideband ($3.39$~GeV$/c^{2}<M_{\pi^{0}\pi^{+}\pi^{-}}<3.45$~GeV$/c^{2}$
and $3.54$~GeV$/c^{2}<M_{\pi^{0}\pi^{+}\pi^{-}}<3.59$~GeV$/c^{2}$).

Exclusive MC samples of $\psi^{\prime}$ decays which have similar
final states are generated to check whether a peak near
$980$~MeV$/c^{2}$ can be produced in the $\pi^{+}\pi^{-}$ mass
spectrum. The main possible backgrounds come from:
$\psi^{\prime}\to\gamma\chi_{c1}\to\gamma\gamma
J/\psi\to\gamma\gamma\pi^{+}\pi^{-}\pi^{0}$;
$\psi^{\prime}\to\pi^{+}\pi^{-}\pi^{0}$;
$\psi^{\prime}\to\gamma\chi_{c1}\to\gamma
a^{\pm}_{0}(980)\pi^{\mp}\to\gamma\eta\pi^{+}\pi^{-}\pi^{0}$ and
$\psi^{\prime}\to\eta\pi^{+}\pi^{-}\pi^{0}$. The $\pi^{+}\pi^{-}$
invariant mass distributions from all these background channels is
shown as the shaded histogram in Fig.~\ref{psip_plot1} (c), and
there is no peak around $980$~MeV$/c^{2}$.

A MC sample of $1.0\times 10^{8}$ inclusive $\psi^{\prime}$ decay
events is used to investigate other possible backgrounds. The shaded
area in Fig.~\ref{psip_plot1} (d) shows the $\pi^{+}\pi^{-}$
invariant mass distribution of events selected from the inclusive MC
sample. In the $980$~MeV$/c^{2}$ region, there is no peaking
background. The $f_{0}(980)$ from other
$\psi^{\prime}\to\gamma\chi_{c1}\to\gamma\pi^{0}
 f_{0}(980)$ processes is much broader than the $a^{0}_{0}(980)\to f_{0}(980)$
mixing signal~\cite{Wu:2008hx} and is estimated from the fit.

\begin{figure}[htbp]
   \centerline{
   \psfig{file=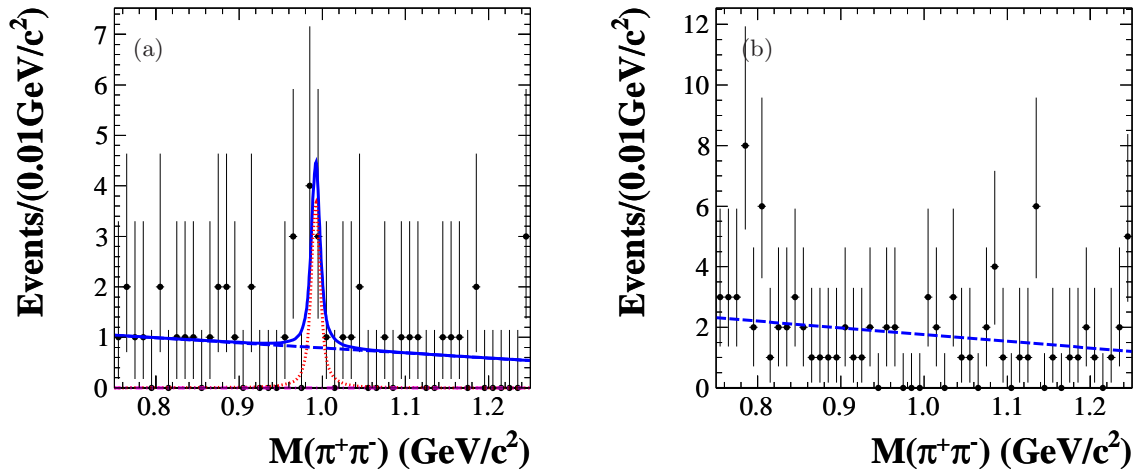,width=16cm, angle=0}
              \put(-400,170){(a)}
              \put(-170,170){(b)}}
   {\caption{
   Results of the simultaneous fit of the $\pi^{+}\pi^{-}$ mass spectra:
   (a) in the $\chi_{c1}$ mass window, (b) in the $\chi_{c1}$ sideband region.
   The solid curve is the result of fit described in the text. The dotted curve shows the mixing signal. The dash-dotted
   curve
   indicates $f_{0}(980)$ from other processes.
   The dashed curves in (a) and (b) denote the
   background polynomial.}
   \label{psip_fit}}
\end{figure}

A simultaneous fit is performed to the $\pi^{+}\pi^{-}$ invariant
mass spectrum in the $\chi_{c1}$ mass window and the $\chi_{c1}$
sideband in a similar manner as in Sect.~\ref{jpsi_channel}. The
$f_{0}(980)$ contribution from other processes is represented by a
Flatt\'{e} formula\footnote{The Flatt\'{e} formula is quoted from
Ref.~\cite{Flatte:1976xu}. The values of coupling constants and mass
used here are quoted from BES~II experiment
results~\cite{Ablikim:2004wn,Wu:2008hx}.}. Figures~\ref{psip_fit}
(a) and (b) show the results of the simultaneous fit.

The fit yields $N(a^{0}_{0}\to f_{0})=6.4\pm3.2$ events for the mixing
signal and $N(f_{0}(980))=0.0\pm8.6$ events for the $f_{0}(980)$
contribution from other processes. Comparing with the fit result
without the mixing signal, the change in $-lnL$ with
$\Delta(d.o.f.)=1$ is $1.79$, corresponding to a statistical
significance of $1.9\sigma$. Comparing with the fit result without the
$f_{0}(980)$ contribution from other processes, the change in $-lnL$
with $\Delta(d.o.f.)=1$ is less than $0.01$, corresponding to a
statistical significance of less than $0.1\sigma$. Using the Bayesian
method, the upper limit for the number of the mixing events is $11.9$,
and the upper limit for the number of the $f_{0}(980)$ events from
other processes is $16.7$ events at the $90\%$ C.L. The results are
listed in Table~\ref{table_psip_fit}.

The branching ratio of the mixing signal
$\psi^{'}\to\gamma\chi_{c1}\to\gamma\pi^{0}
a^{0}_{0}(980)\to\gamma\pi^{0}
f_{0}(980)\to\gamma\pi^{0}\pi^{+}\pi^{-}$ is calculated as:
\begin{eqnarray*}
Br(\psi^{'}\to\gamma\chi_{c1}\to\gamma\pi^{0} a^{0}_{0}(980)
\to\gamma\pi^{0} f_{0}(980)\to\gamma\pi^{0}\pi^{+}\pi^{-})= \frac
{N(a^{0}_{0}\to f_{0})} {\varepsilon_{af} \cdot N_{\psi^{\prime}}
\cdot Br(\pi^{0}\to\gamma\gamma)},
\end{eqnarray*}
where $N_{\psi^{\prime}}$ is the total number of $\psi^{\prime}$
events and $\varepsilon_{af}=(22.3\pm0.2)\%$ is the efficiency for
the mixing signal $\psi^{\prime}\to\gamma\chi_{c1}\to\gamma\pi^{0}
a^{0}_{0}(980)\to\gamma\pi^{0}
f_{0}(980)\to\gamma\pi^{0}\pi^{+}\pi^{-}$. The branching ratio is
then determined to be $(2.7\pm1.4)\times10^{-7}$, where the error is
statistical only.

The total branching ratio of
$\psi^{'}\to\gamma\chi_{c1}\to\gamma\pi^{0}\pi^{+}\pi^{-}$ is
calculated as:
\begin{eqnarray*}
Br(\psi^{'}\to\gamma\chi_{c1}\to\gamma\pi^{0} a^{0}_{0}(980)
\to\gamma\pi^{0} f_{0}(980)\to\gamma\pi^{0}\pi^{+}\pi^{-})=\frac
{N(a^{0}_{0}\to f_{0})/\varepsilon_{af}+N(f_{0})/\varepsilon_{f} }
{N_{\psi^{\prime}} \cdot Br(\pi^{0}\to\gamma\gamma)},
\end{eqnarray*}
where $\varepsilon_{f}=(20.5\pm0.2)\%$ is the efficiency for the
underlying process $\psi^{\prime}\to\gamma\pi^{0}
f_{0}(980)\to\gamma\pi^{0}\pi^{+}\pi^{-}$. The branching ratio is
then determined to be $(2.7\pm4.2)\times10^{-7}$, where the error is
statistical only.

\begin {table}[htp]
\begin {center}
\caption {Results of fits to the $\pi^{+}\pi^{-}$ invariant mass
spectrum in the $\chi_{c1}$ region. The fit is described in the text and
the yield error is only statistical. N($a^{0}_{0}\to f_{0}$) is the
number of mixing events $a^{0}_{0}(980)\to f_{0}(980)$.
S($a^{0}_{0}\to f_{0}$) is the significance of the mixing signal
$a^{0}_{0}(980)\to f_{0}(980)$. N($f_{0}(980)$) is the number of
$f_{0}(980)$ events from other processes. S($f_{0}(980)$) is the
significance of the $f_{0}(980)$ contribution from other processes.}
\label{table_psip_fit} \vspace{0.2cm}
\begin {tabular}{c|c|c|c|c} \hline
    Mode                                &N($a^{0}_{0}\to f_{0}$)        &S($a^{0}_{0}\to f_{0}$)        &N($f_{0}(980)$)         &S($f_{0}(980)$)     \\\hline  \hline
    mixing + $f_{0}(980)$ + 2nd poly.   &$6.4\pm3.2(<11.9)$             &$1.9\sigma$                    &$0.0 \pm8.6(<16.7)$      &$<0.1\sigma$        \\  \hline
    mixing + 2nd poly.                  &$6.4\pm3.2(<12.1)$             &$3.0\sigma$                    &$-$                     &$-$                 \\  \hline
    $f_{0}(980)$ + 2nd poly.            &$-$                            &$-$                            &$12.8\pm 6.7(<23.6)$    &$2.3\sigma$         \\  \hline
    \end {tabular}
\end {center}
\end {table}

If we fit the $\pi^{+}\pi^{-}$ invariant mass spectrum only with the
mixing signal plus a 2nd order polynomial background, the fit yields
$6.4\pm3.2$ events for the mixing signal. Comparing with the fit
result with only the 2nd order polynomial, the change in $-lnL$ with
$\Delta(d.o.f.)=1$ is $4.41$, corresponding to a statistical
significance of $3.0\sigma$. The upper limit at the $90\%$ C.L. is
$12.1$ events.

If we assume there is no mixing and fit the $\pi^{+}\pi^{-}$
invariant mass spectrum only with the $f_{0}(980)$ contribution from
other processes plus a 2nd order polynomial, the fit yields
$12.8\pm6.7$ events for the $f_{0}(980)$ contribution from other
processes. Comparing with the fit result with only the 2nd order
polynomial, the change in $-lnL$ with $\Delta(d.o.f.)=1$ is $2.62$,
corresponding to a statistical significance of $2.3\sigma$. The
upper limit at the $90\%$ C.L. is $23.6$ events. The fit results are
listed in Table~\ref{table_psip_fit}. The total branching ratio of
$\psi^{'}\to\gamma\chi_{c1}\to\gamma\pi^{0}
f_{0}(980)\to\gamma\pi^{0}\pi^{+}\pi^{-}$ is calculated to be
$(6.0\pm3.1)\times10^{-7}$, where the error is statistical only.

\section{Discussion}\label{Discussion}
Various models for the $a^{0}_{0}(980)$ and the
$f_{0}(980)$~\cite{Achasov:1987ts,Achasov:1997ih,
Weinstein:1983gd,Weinstein:1990gu, SI_conf1995} give different
predictions for their coupling constants and masses; these have been
measured by several
experiments~\cite{Achasov:2000ym,Achasov:2000ku,Aloisio:2002bsa,
Aloisio:2002bt,Teige:1996fi,Bugg:1994mg}. From these theoretical and
experimental values of the resonance parameters, predictions for
$\xi_{af}$ and $\xi_{fa}$ are calculated~\cite{Wu:2007jh,Wu:2008hx}.
Using the parameter sets listed in Table~\ref{table_shape}, the line
shapes of the $a^{0}_{0}(980)\to f_{0}(980)$ and $f_{0}(980)\to
a^{0}_{0}(980)$ mixing signals can be determined from MC simulation
and the underlying $f_{0}(980)$ ($a^{0}_{0}(980)$) shapes can be
parameterized accordingly. Table~\ref{table_shape} shows the fitting
results obtained using a similar fitting procedure as described in
Sects.~\ref{jpsi_channel} and ~\ref{psip_channel}. The fitting
results are consistent within the statistical error.

\begin {table}[htp]
\begin {center}
\caption{Fitting results with various theoretical and experimental
values of the resonance parameters.} \label{table_shape}
\vspace{0.2cm}
\begin {tabular}{c|c|c|c|c} \hline
    Mixing shape                                                        &N($f_{0}\to a^{0}_{0}$)          &S($f_{0}\to a^{0}_{0}$)          &N($a^{0}_{0}\to f_{0}$)          &S($a^{0}_{0}\to f_{0}$)        \\\hline\hline
    $q\bar{q}$~\cite{Achasov:1987ts}                                    &$19.8\pm 8.6 (<31.8)$            &$2.4\sigma$                              &$5.9\pm 2.8(<10.9)$              &$2.0\sigma$                    \\
    $q^{2}\bar{q}^{2}$~\cite{Achasov:1987ts}                            &$19.4\pm 8.5 (<31.5)$            &$2.4\sigma$                              &$6.3\pm 3.0(<11.5)$              &$2.1\sigma$                    \\
    $K\bar{K}$~\cite{Achasov:1997ih,Weinstein:1983gd,Weinstein:1990gu}  &$14.5\pm 10.8(<28.3)$            &$1.3\sigma$                              &$5.8\pm 2.7(<10.5)$              &$1.6\sigma$                    \\
    $q\bar{q}g$~\cite{SI_conf1995}                                      &$25.4\pm 9.7 (<38.2)$            &$2.9\sigma$                              &$6.6\pm 3.2(<12.2)$              &$2.6\sigma$                    \\  \hline
    SND~\cite{Achasov:2000ym,Achasov:2000ku}                            &$21.7\pm 9.3 (<33.1)$            &$2.5\sigma$                              &$6.0\pm 2.9(<11.1)$              &$2.0\sigma$                    \\
    KLOE~\cite{Aloisio:2002bsa,Aloisio:2002bt}                          &$23.3\pm 8.0 (<34.9)$            &$3.3\sigma$                              &$6.3\pm 3.0(<11.6)$              &$2.0\sigma$                    \\
    BNL~\cite{Teige:1996fi}                                             &$28.7\pm 6.8 (<38.7)$            &$4.1\sigma$                              &$6.4\pm 3.0(<11.8)$              &$2.4\sigma$                    \\
    CB~\cite{Bugg:1994mg}                                               &$27.1\pm 8.4 (<37.8)$            &$3.7\sigma$                              &$6.4\pm 3.1(<11.8)$              &$2.2\sigma$                    \\  \hline
\end {tabular}
\\
\end {center}
\end {table}

\section{Systematic uncertainties}\label{Syserr}

The systematic uncertainties on the branching ratios are summarized
in Table~\ref{table_err}.

The systematic uncertainty associated with the tracking efficiency
has been studied with control samples such as
$J/\psi\to\rho\pi$, $J/\psi\to p\bar{p}\pi^{+}\pi^{-}$, $J/\psi\to
K^{*}K\to K_{S}K\pi$. The difference of the tracking efficiencies
between data and MC simulation is $2\%$ per charged track.

The uncertainties due to PID of $\pi$ and $K$ are determined from
studies of control samples such as $J/\psi\to\rho\pi$, $J/\psi\to
p\bar{p}\pi^{+}\pi^{-}$, $J/\psi\to K^{*}K\to K^{+}K^{-}\pi^{0}$.
The difference of the PID efficiency between data and MC is $2\%$
per track.

The uncertainty due to photon detection and photon conversion is
$1\%$ per photon. This is determined from studies of the photon
detection efficiency in control samples such as
$J/\psi\to\rho^{0}\pi^{0}$ and a study of photon conversion via
$e^{+}e^{-}\to\gamma\gamma$.

The uncertainty due to the $\pi^{0}$ selection is determined from a high
purity control sample of $J/\psi\to\pi^{+}\pi^{-}\pi^{0}$ decays.
The $\pi^{0}$ selection efficiency is obtained from the change in
the $\pi^{0}$ yield in the $\pi^{+}\pi^{-}$ recoiling mass spectrum
with or without the $\pi^{0}$ selection requirement. The difference
of the $\pi^{0}$ reconstruction efficiency between data and MC
simulation gives an uncertainty of $2.0\%$ per $\pi^{0}$. The
uncertainty from the $\eta$ selection is $2.0\%$ per $\eta$, which
is determined in a similar way from a control sample of $J/\psi\to
p\bar{p}\eta$.

The uncertainty of the kinematic fit for $J/\psi\to\phi
f_{0}(980)\to\phi a^{0}_{0}(980)\to\phi\eta\pi^{0}$ is estimated
from $J/\psi\to\omega\eta\to\pi^{+}\pi^{-}\pi^{0}\eta$
($\eta,\pi^{0}\to\gamma\gamma$). The efficiency is obtained from the
change in the yield of $\omega$ signal by a fit to the
$\pi^{+}\pi^{-}\pi^{0}$ mass distribution with or without the
requirement of the kinematic fit. The systematic uncertainty is
determined to be $0.9\%$. The uncertainty of the kinematic fit for
$\psi^{\prime}\to \gamma\chi_{c1}\to \gamma\pi^{0}
a^{0}_{0}(980)\to\gamma\pi^{0} f_{0}(980)\to
\gamma\pi^{0}\pi^{+}\pi^{-}$ is estimated to be $1.7\%$ from
$\psi^{\prime}\to\pi^{+}\pi^{-} J/\psi$, $J/\psi\to\gamma\eta$.

The branching ratios for the $\eta\to\gamma\gamma$,
$\pi^{0}\to\gamma\gamma$ and $\phi\to K^{+}K^{-}$ decays are taken
from the PDG~\cite{PDG}. The uncertainty on these branching ratios
are taken as a systematic uncertainty in our measurements.

The total number of $J/\psi$ events is $(2.252\pm0.028)\times 10^{8}$,
determined from inclusive $J/\psi$ hadronic decays~\cite{yanghx}, and the
total number of $\psi^{\prime}$ events is $(1.06\pm0.04)\times 10^{8}$,
determined from inclusive $\psi^{\prime}$ hadronic events~\cite{Ablikim:2010zn}. The
uncertainty on the number of $J/\psi$ events is $1.3\%$, and the
uncertainty on the number of $\psi^{\prime}$ events is $4\%$.

To estimate the systematic uncertainties due to the fit procedure,
we repeat the fit with appropriate modifications to estimate the
systematic uncertainties. The largest difference of the yield of
each sources with respect to the values derived from the standard
fit is considered as a systematic error. We change the sideband
range and the order of polynomial to estimate the uncertainty from
the background shape. A series of fits using different fitting
ranges is performed and the largest change of the branching ratios
is assigned as a systematic uncertainty.

The total systematic uncertainties for the branching ratio
measurements are obtained by adding up the contributions from all
the systematic sources in quadrature.

The uncertainty due to the parameterization of the mixing signal
line shape and the underlying $a^{0}_{0}(980)$ ($f_{0}(980)$) is
kept separate and quoted as a second systematic uncertainty. The
uncertainty is obtained by comparing the results with the parameter
sets in Table~\ref{table_shape} with the standard fit. We take this
difference as a conservative estimate of the uncertainty and assign
an uncertainty of 43.8\% for the $a^{0}_{0}(980)\to f_{0}(980)$
mixing measurement and an uncertainty of 9.4\% for the
$f_{0}(980)\to a^{0}_{0}(980)$ mixing measurement. For the total
branching ratio measurement of $J/\psi\to\phi
a^{0}_{0}(980)\to\phi\eta\pi^{0}$, the uncertainty is assigned to be
38.3\% and for the total branching ratio measurement of
$\psi^{\prime}\to\gamma\chi_{c1}\to\gamma\pi^{0}
f_{0}(980)\to\gamma\pi^{0}\pi^{+}\pi^{-}$, the uncertainty is
assigned to be 9.4\%. If we assume there is no mixing, the
uncertainties of the total branching ratio measurements
are assigned to be 41.4\% and 42.2\%, respectively.

\begin {table}[htp]
\begin {center}
\caption {Summary of systematic errors for the branching ratio  
measurements and upper limits determination.} \label{table_err}
\vspace{0.2cm}
\begin {tabular}{c|c|c|c|c} \hline
    $J/\psi\to\phi
a^{0}_{0}(980)\to\phi\eta\pi^0$            & mixing Br & upper limit
of mixing Br & total Br& total Br(no mixing)
\\\hline \hline
    Charged tracks      &$4.0\%$                        &$4.0\%$               &$4.0\%$  &$4.0\%$                \\  \hline
    Photon detection    &$4.0\%$                        &$4.0\%$               &$4.0\%$    &$4.0\%$            \\  \hline
    PID                 &$4.0\%$                        &$4.0\%$               &$4.0\%$     &$4.0\%$          \\  \hline
    $\eta$ construction &$2.0\%$                        &$2.0\%$               &$2.0\%$      &$2.0\%$          \\  \hline
    $\pi^{0}$ construction  &$2.0\%$                    &$2.0\%$               &$2.0\%$      &$2.0\%$            \\  \hline
    Kinematic fit       &$0.9\%$                        &$0.9\%$                 &$0.9\%$      &$0.9\%$          \\  \hline
    Intermediate decay  &$0.54\%$                       &$0.54\%$                &$0.54\%$   &$0.54\%$           \\  \hline
    Normalization       &$1.3\%$                        &$1.3\%$                  &$1.3\%$    &$1.3\%$        \\  \hline
    Background shape    &$6.6\%$                        &$-$                       &$28.7\%$   &$4.9\%$            \\  \hline
    Fitting range       &$6.2\%$                        &$-$                       &$14.5\%$    &$12.9\%$        \\  \hline
    Total          &$11.9\%$                       &$7.7\%$                 &$33.1\%$   &$15.8\%$          \\  \hline
\end {tabular}

\begin {tabular}{c|c|c|c|c} \hline
    $\psi^{\prime}\to\gamma\chi_{c1}\to\gamma\pi^{0}
f_{0}(980)$& mixing Br & upper limit of mixing Br & total Br & total
Br (no mixing)\\
$\to\gamma\pi^{0}\pi^{+}\pi^{-}$&&&&\\\hline \hline
    Charged tracks                &$4.0\%$               &$4.0\%$            &$4.0\%$   &$4.0\%$     \\  \hline
    Photon detection            &$3.0\%$               &$3.0\%$           &$3.0\%$     &$3.0\%$  \\  \hline
    PID                         &$4.0\%$               &$4.0\%$             &$4.0\%$    &$4.0\%$     \\  \hline
    $\pi^{0}$ construction           &$2.0\%$               &$2.0\%$             &$2.0\%$    &$2.0\%$  \\  \hline
    Kinematic fit         &$1.7\%$               &$1.7\%$             &$1.7\%$     &$1.7\%$    \\  \hline
    Intermediate decay      &$0.034\%$             &$0.034\%$          &$0.034\%$    &$0.034\%$   \\  \hline
    Normalization             &$4.0\%$               &$4.0\%$               &$4.0\%$   &$4.0\%$   \\  \hline
    Background shape            &$23.4\%$              &$-$             &$128.1\%$     &$48.4\%$      \\  \hline
    Fitting range              &$6.3\%$               &$-$                 &$32.8\%$  &$17.2\%$ \\  \hline
    Total                 &$25.5\%$              &$8.0\%$           &$132.5\%$  &$52.0\%$    \\  \hline
\end {tabular}

\end {center}
\end {table}

\section{Summary}\label{Summary}

Based on $(2.252\pm0.028)\times 10^{8}$ $J/\psi$ events and
$(1.06\pm0.04)\times 10^{8}$ $\psi^{\prime}$ events, the mixing
branching ratios are measured to be: $Br(J/\psi\to\phi
f_{0}(980)\to\phi a^{0}_{0}(980)\to\phi\eta\pi^{0})$ = $(3.3\pm1.1$
(stat.)$\pm0.4$ (sys.)$\pm1.4$ (para.)$)\times 10^{-6}$,
$Br(\psi^{\prime}\to\gamma\chi_{c1}\to\gamma\pi^{0}
a^{0}_{0}(980)\to\gamma\pi^{0}
f_{0}(980)\to\gamma\pi^{0}\pi^{+}\pi^{-})$ = $(2.7\pm1.4$
(stat.)$\pm0.7$ (sys.)$\pm0.3$ (para.)$)\times 10^{-7}$, where the
uncertainties are statistical, systematics due to this measurement,
and systematics due to the parameterization, respectively.

The total branching ratio of $J/\psi\to\phi
a^{0}_{0}(980)\to\phi\eta\pi^{0}$ is measured to be $(5.0\pm2.7$
(stat.)$\pm1.7$ (sys.)$\pm1.9$ (para.)$)\times 10^{-6}$ and the
total branching ratio of
$\psi^{\prime}\to\gamma\chi_{c1}\to\gamma\pi^{0}
f_{0}(980)\to\gamma\pi^{0}\pi^{+}\pi^{-}$ is measured to be
$(2.7\pm4.2$ (stat.)$\pm3.6$ (sys.)$\pm0.3$ (para.)$)\times
10^{-7}$. If we assume there's no mixing, the total branching ratios
are measured to be $(9.7\pm2.2$ (stat.)$\pm1.5$ (sys.)$\pm4.0$
(para.)$)\times 10^{-6}$ and $(6.0\pm3.1$ (stat.)$\pm3.1$
(sys.)$\pm2.5$ (para.)$)\times 10^{-7}$, respectively.


When determining the upper limit of the number of $J/\psi\to\phi
f_{0}(980)\to\phi a^{0}_{0}(980)\to\phi\eta\pi^{0}$ events, the
uncertainties due to the fit range, background shape, the
parameterization of mixing signal line shape and the underlying
$a^{0}_{0}(980)$ are considered. Using the Bayesian method,
different upper limits at the $90\%$ C.L. are determined by varying
the fit range, the background shape and the parameterization of the
mixing signal line shape and the underlying $a^{0}_{0}(980)$ in
Table~\ref{table_shape}. The upper limit for the mixing signal is
taken to be the largest of them: $N^{U.L.}_{fa} = 39.7$ events. A
conservative estimate of the upper limit of the branching ratio is
determined by lowering the efficiency by one standard deviation.

The upper limit on the branching ratio at the $90\%$ C.L. is calculated as :
\begin{eqnarray*}
Br(J/\psi\to\phi
f_{0}(980)\to\phi a^{0}_{0}(980)\to\phi\eta\pi^{0})~~~~~~~~~~~~~~~~~~~~~~~~~~~~~~~~\\
< \frac {N^{U.L.}_{fa}} {(\varepsilon_{fa} -\sigma^{U.L.}_{fa})
\cdot N_{J/\psi} \cdot Br(\phi\to K^{+}K^{-}) \cdot
Br(\eta\to\gamma\gamma) \cdot Br(\pi^{0}\to\gamma\gamma)}= 5.4\times
10^{-6}.
\end{eqnarray*}

Similarly, considering the uncertainties due to fit range,
background shape, the parameterization of mixing signal line shape
and the underlying $f_{0}(980)$ , the upper limit number of the
mixing signal $\psi^{\prime}\to\gamma\chi_{c1}\to\gamma\pi^{0}
a^{0}_{0}(980)\to\gamma\pi^{0}
f_{0}(980)\to\gamma\pi^{0}\pi^{+}\pi^{-}$ is determined to be:
$N^{U.L}_{af} = 13.0$ events.

The upper limit on the branching ratio at the $90\%$ C.L. is calculated as :
\begin{eqnarray*}
Br(\psi^{\prime}\to\gamma\chi_{c1}\to\gamma\pi^{0}
a^{0}_{0}(980)\to\gamma\pi^{0}
f_{0}(980)\to\gamma\pi^{0}\pi^{+}\pi^{-})\\
< \frac {N^{U.L.}_{af}} {(\varepsilon_{af} -\sigma^{U.L.}_{af})
\cdot N_{\psi^{\prime}}
\cdot Br(\pi^{0}\to\gamma\gamma)} = 6.0\times 10^{-7}.\\
\end{eqnarray*}

The mixing intensity $\xi_{fa}$ for the $f_{0}(980)\to
a^{0}_{0}(980)$ transition is calculated to be:
\begin{eqnarray*}\label{eq_xifa}
\xi_{fa} = \frac {Br(J/\psi\to\phi f_{0}(980)\to\phi
a^{0}_{0}(980)\to\phi\eta\pi^{0})} {Br(J/\psi\to\phi
f_{0}(980)\to\phi\pi\pi)\textsuperscript
{\cite{Ablikim:2004wn}}}\\
=(0.60\pm0.20 (stat.)\pm0.12 (sys.)\pm0.26 (para.)\%,
\end{eqnarray*}
where the uncertainties are statistical, systematics due to this
measurement, and the parameterization, respectively. The uncertainty
from $Br(J/\psi\to\phi f_{0}(980)\to\phi\pi\pi)$ is included as a
part of the systematic error. The upper limit of the mixing
intensity $\xi_{fa}$ at the $90\%$ C.L. is $1.1\%$.

The mixing intensity $\xi_{af}$ for the $a^{0}_{0}(980)\to
f_{0}(980)$ transition is calculated to be:
\begin{eqnarray*}\label{eq_xiaf}
\xi_{af} = \frac {Br(\psi^{\prime}\to\gamma\chi_{c1}\to\gamma\pi^{0}
a^{0}_{0}(980)\to\gamma\pi^{0}
f_{0}(980)\to\gamma\pi^{0}\pi^{+}\pi^{-})}
{Br(\psi^{\prime}\to\gamma\chi_{c1}\to\gamma\pi^{0}
a^{0}_{0}(980)\to\gamma\pi^{0}\pi^{0}\eta)\textsuperscript
{\cite{PDG}}}\\
=(0.31\pm0.16 (stat.)\pm0.14 (sys.)\pm0.03 (para.))\%.
\end{eqnarray*}
The uncertainty from
$Br(\psi^{\prime}\to\gamma\chi_{c1}\to\gamma\pi^{0}
a^{0}_{0}(980)\to\gamma\pi^{0}\pi^{0}\eta)$ is included as a part of
the systematic error. The upper limit of the mixing intensity
$\xi_{af}$ at the $90\%$ C.L. is $1.0\%$.

\begin{figure}[htbp]
   \centerline{
   \psfig{file=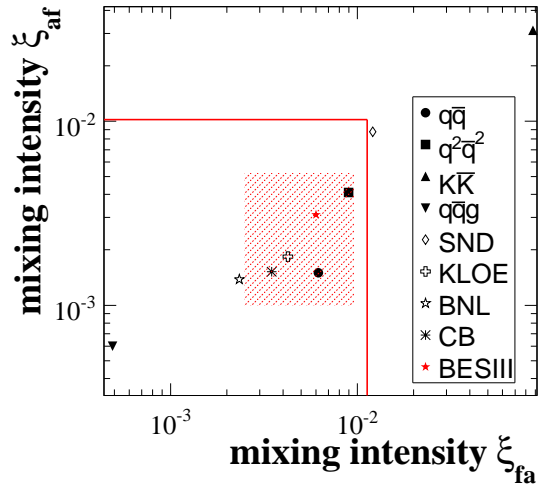,width=8.0cm, angle=0}}
   {\caption{Mixing intensities of $f_{0}(980) \to a^{0}_{0}(980)$
   and $a^{0}_{0}(980) \to f_{0}(980)$.The dots are predictions for the
   mixing intensities with theoretical and experimental values of the
   parameters~\cite{Wu:2008hx}. The shaded region is our results with
   statistical errors and systematics due to this
measurement and the parameterization. The solid lines mark the upper
limits.}
   \label{intensity}}
\end{figure}

The calculated mixing intensities~\cite{Wu:2008hx} with the
resonance parameters from various
models~\cite{Achasov:1987ts,Achasov:1997ih,Weinstein:1983gd,Weinstein:1990gu,
SI_conf1995} and experimental
measurements~\cite{Achasov:2000ym,Achasov:2000ku,
Aloisio:2002bsa,Aloisio:2002bt,Teige:1996fi,Bugg:1994mg} are
compared with our results in Fig.~\ref{intensity}. This result will
be very useful in pinning down the resonance parameters of
$a^{0}_{0}(980)$ and $f_{0}(980)$.



\section{Acknowledgments}

The BES~III collaboration thanks the staff of BEPC~II and the
computing center for their hard efforts. This work is supported in
part by the Ministry of Science and Technology of China under
Contract No. 2009CB825200; National Natural Science Foundation of
China (NSFC) under Contracts Nos. 10625524, 10821063, 10825524,
10835001, 10875113, 10935007, 10979008, 10979038, 11005109,
11079030; the Chinese Academy of Sciences (CAS) Large-Scale
Scientific Facility Program; CAS under Contracts Nos. KJCX2-YW-N29,
KJCX2-YW-N45; 100 Talents Program of CAS; Research Fund for the
Doctoral Program of Higher Education of China under Contract No.
20093402120022; Istituto Nazionale di Fisica Nucleare, Italy;
Russian Foundation for Basic Research under Contracts Nos.
08-02-92221, 08-02-92200-NSFC-a; Siberian Branch of Russian Academy
of Science, joint project No 32 with CAS; U. S. Department of Energy
under Contracts Nos. DE-FG02-04ER41291, DE-FG02-91ER40682,
DE-FG02-94ER40823; University of Groningen (RuG) and the
Helmholtzzentrum fuer Schwerionenforschung GmbH (GSI), Darmstadt;
WCU Program of National Research Foundation of Korea under Contract
No. R32-2008-000-10155-0.



\begin{thebibliography}{99}

\bibitem{Jaffe:1976ig}
  R.~L.~Jaffe,
  Phys.\ Rev.\  D {\bf 15}, 267 (1977).


\bibitem{Achasov:1987ts}
  N.~N.~Achasov and V.~N.~Ivanchenko,
  Nucl.\ Phys.\  B {\bf 315}, 465 (1989).


\bibitem{Achasov:1997ih}
  N.~N.~Achasov and V.~V.~Gubin,
  Phys.\ Rev.\  D {\bf 56}, 4084 (1997).


\bibitem{Weinstein:1983gd}
  J.~D.~Weinstein and N.~Isgur,
  Phys.\ Rev.\  D {\bf 27}, 588 (1983).


\bibitem{Weinstein:1990gu}
  J.~D.~Weinstein and N.~Isgur,
  Phys.\ Rev.\  D {\bf 41}, 2236 (1990).


\bibitem{SI_conf1995} 
S.~Ishida {\it et al.}, in the 6th International Converence on Hadron
Spectroscopy, Manchester, UK, 1995, p. 454.


\bibitem{Achasov:1979xc}
  N.~N.~Achasov, S.~A.~Devyanin and G.~N.~Shestakov,
  Phys.\ Lett.\  B {\bf 88}, 367 (1979).


\bibitem{Hanhart:2007bd}
  C.~Hanhart, B.~Kubis and J.~R.~Pelaez,
  Phys.\ Rev.\  D {\bf 76}, 074028 (2007).


\bibitem{Achasov:2002hg}
  N.~N.~Achasov and A.~V.~Kiselev,
  Phys.\ Lett.\  B {\bf 534}, 83 (2002).


\bibitem{Kerbikov:2000pu}
  B.~Kerbikov and F.~Tabakin,
  Phys.\ Rev.\  C {\bf 62}, 064601 (2000).


\bibitem{Achasov:2003se}
  N.~N.~Achasov and G.~N.~Shestakov,
  Phys.\ Rev.\ Lett.\  {\bf 92}, 182001 (2004).


\bibitem{Close:2000ah}
  F.~E.~Close and A.~Kirk,
  Phys.\ Lett.\  B {\bf 489}, 24 (2000).


\bibitem{Wu:2007jh}
  J.~J.~Wu, Q.~Zhao and B.~S.~Zou,
  Phys.\ Rev.\  D {\bf 75}, 114012 (2007).


\bibitem{Wu:2008hx}
  J.~J.~Wu and B.~S.~Zou,
  Phys.\ Rev.\  D {\bf 78}, 074017 (2008).


\bibitem{yanghx} 
  M. Ablikim {\it et al.}, Measurement of the Matrix Element for the Decay
  $\eta\prime \to \eta\pi^{+}\pi^{-}$, submitted to Phys. Rev. D.
  [arXiv:1012.1117 [hep-ex]].


\bibitem{Ablikim:2010zn}
  M.~Ablikim {\it et al.},
  Phys.\ Rev.\  D {\bf 81}, 052005 (2010).


\bibitem{PDG}
K.~Nakamura {\it et al.}  (Particle Data Group), J. of Phys. G {\bf 37}, 075021 (2010).


\bibitem{Flatte:1976xu}
  S.~M.~Flatte,
  Phys.\ Lett.\  B {\bf 63}, 224 (1976).


\bibitem{Bugg:1994mg}
  D.~V.~Bugg, V.~V.~Anisovich, A.~Sarantsev and B.~S.~Zou,
  Phys.\ Rev.\  D {\bf 50}, 4412 (1994).


\bibitem{Ablikim:2004wn}
  M.~Ablikim {\it et al.}  [BES Collaboration],
  Phys.\ Lett.\  B {\bf 607}, 243 (2005).


\bibitem{Achasov:2000ym}
  M.~N.~Achasov {\it et al.},
  Phys.\ Lett.\  B {\bf 485}, 349 (2000).


\bibitem{Achasov:2000ku}
  M.~N.~Achasov {\it et al.},
  Phys.\ Lett.\  B {\bf 479}, 53 (2000).


\bibitem{Aloisio:2002bsa}
  A.~Aloisio {\it et al.}  [KLOE Collaboration],
  Phys.\ Lett.\  B {\bf 536}, 209 (2002).


\bibitem{Aloisio:2002bt}
  A.~Aloisio {\it et al.}  [KLOE Collaboration],
  Phys.\ Lett.\  B {\bf 537}, 21 (2002).


\bibitem{Teige:1996fi}
  S.~Teige {\it et al.}  [E852 Collaboration],
  Phys.\ Rev.\  D {\bf 59}, 012001 (1999).



\end{thebibliography}
\end{document}